\documentclass[12pt]{article}
\usepackage{epsfig,url}
\usepackage{lineno}
\newcommand\eps{\varepsilon}

\begin{document}
%\linenumbers
\title{Beam parameters of a M\"obius ring}
\author{V. Ziemann, FREIA, Uppsala University}
\date{\today}
\maketitle
%%%%%%%%%%%%%%%%%%%%%%%%%%%%%%%%%%%%%%%%%%%%%%%%%%%%%%%%%%%%%%%%%%%%%%%
\begin{abstract}\noindent
  We describe a racetrack-shaped ring in which the cross-plane coupling can be
  continuously varied between an uncoupled configuration to a M\"obius configuration,
  where the transverse planes change role after one turn. Tunes, emittances, and
  other beam parameters that depend on the emission of synchrotron radiation are
  calculated as a function of the coupling.
\end{abstract}
%
%%%%%%%%%%%%%%%%%%%%%%%%%%%%%%%%%%%%%%%%%%%%%%%%%%%%%%%%%%%%%%%%%%%%%%%
%
\section{Introduction}
In order to explore a new algorithm~\cite{SZ} to calculate the radiation
integrals~\cite{HELM} we designed an accelerator lattice for a storage ring
in which the transverse coupling can be continuously varied; all the way
from the uncoupled lattice to a M\"obius configuration~\cite{TALMAN} and
even beyond. Then we calculate the radiation integrals for this strongly
coupled lattice and derive beam parameters such as the emittances.
\par
%..............................................
\begin{figure}[tb]
\begin{center}
\includegraphics[width=0.41\textwidth]{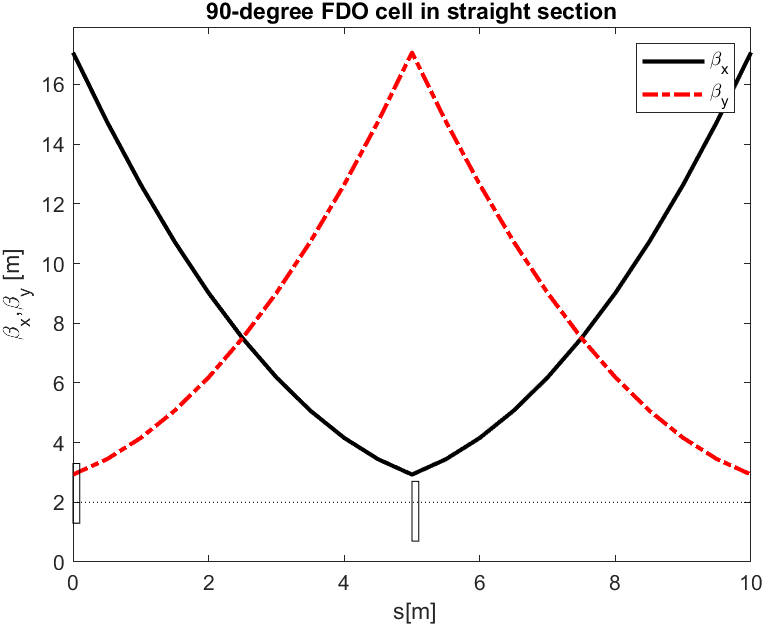}
\includegraphics[width=0.47\textwidth]{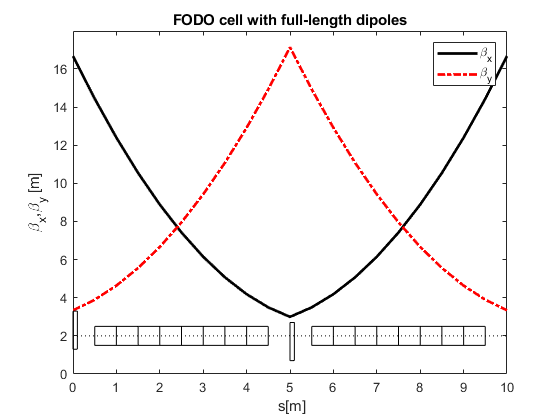}
\end{center}
\caption{\label{fig:cell}The beta function of the basic 90-degree cell (left)
  and a cell with 4\,m-long dipoles used in the arcs (right).}
\end{figure}
% ..............................................
All MATLAB scripts used to design this ring are available from~\cite{SUPMAT}.
The software is based on~\cite{VZAP} but is adapted to handle coupling and
dispersion effects simultaneously.
In order to keep the design conceptually simple, we use thin-lens quadrupoles,
both for the upright and the skew quadrupoles and base our design on the $90^o$
FODO cell, shown on the left-hand side in Figure~\ref{fig:cell}. This cell
has a length of $2L=10\,$m, where $L=5\,$m is the length of the drift space
between the quadrupoles, which have a focal length of $L/\sqrt{2}$, giving the
phase-advance of 90\,degrees per cell in both planes. The beta functions
assume their maximum and minimum value of 17.7 and 2.9\,m, respectively.
The value in the middle of the straight section is $3L/2=7.5\,$m. 
\par
In the arcs we insert 4\,m long dipoles with a deflection angle of 2\,degrees
in each of the straight sections between the quadrupoles. The right-hand side
in Figure~\ref{fig:cell} illustrates this. In order to avoid equal integer tunes
in both planes we adjust the quadrupoles in these cells to give phase advances
of $0.25\times 360^o$ and $0.22\times 360^o$ in the horizontal and vertical
plane, respectively.
\section{Arcs}
%
%..............................................
\begin{figure}[tb]
\begin{center}
\includegraphics[width=0.47\textwidth]{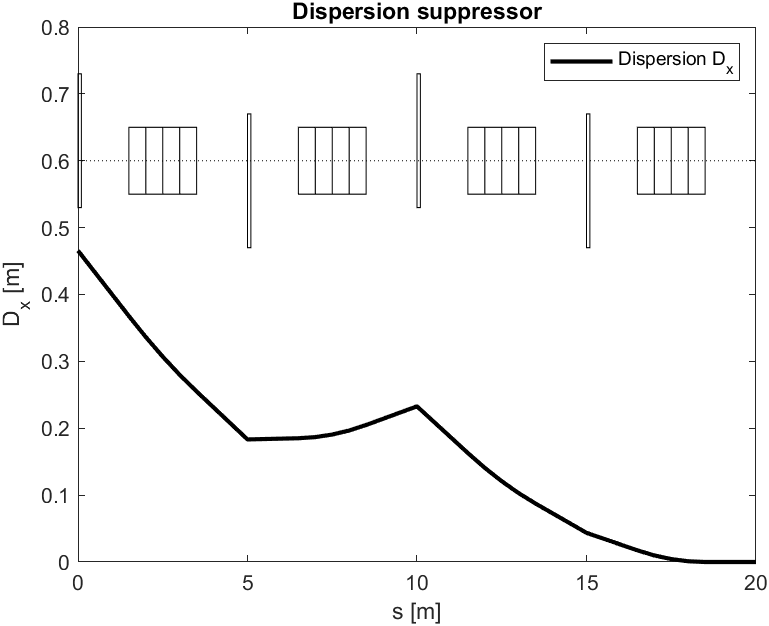}
\includegraphics[width=0.47\textwidth]{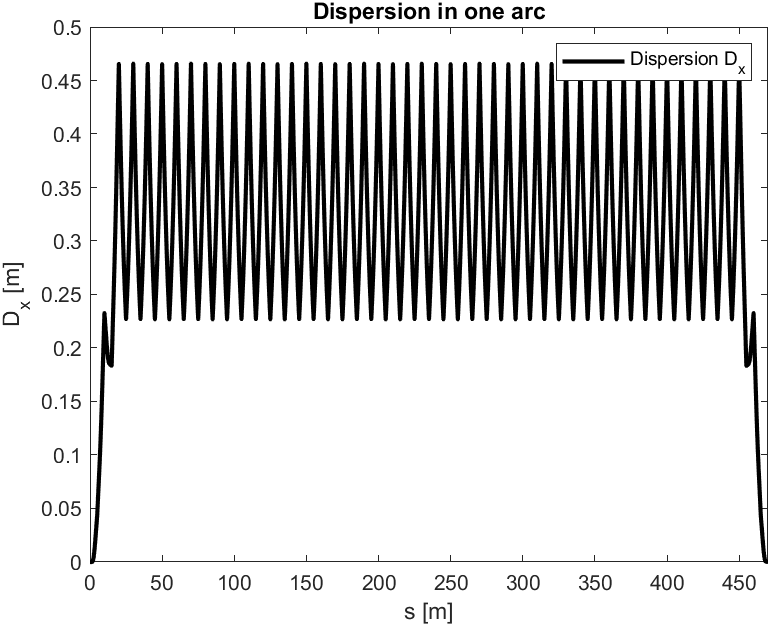}
\end{center}
\caption{\label{fig:disp}The horizontal dispersion function $D_x$ in the
  dispersion suppressor (left) and the dispersion across one arc.}
\end{figure}
%..............................................
Each of the two arcs consist of 43 standard cells shown on the right-hand side in
Figure~\ref{fig:cell}
and two cells with half-length dipoles each on either end. These additional cells create an
interference pattern in the periodically oscillating dispersion that cancels the dispersion
almost perfectly. We adjust the two QF and two QD in the dispersion suppressor to
perfectly cancel the dispersion and its derivative. This slightly changes the beta
functions and the phase advance across the arc. We will correct that later in the straight
sections with additional matching quadrupoles. The left-hand plot in Figure~\ref{fig:disp}
shows the dispersion suppressor and the plot on the right-hand side shows the dispersion
across one arc. We observe that the dispersion
suppressors at the entrance and exit of the arc---both having the same quadrupole
excitations---nicely match the dispersion.
\section{Straight sections}
%
%..............................................
\begin{figure}[tb]
\begin{center}
\includegraphics[width=0.7\textwidth]{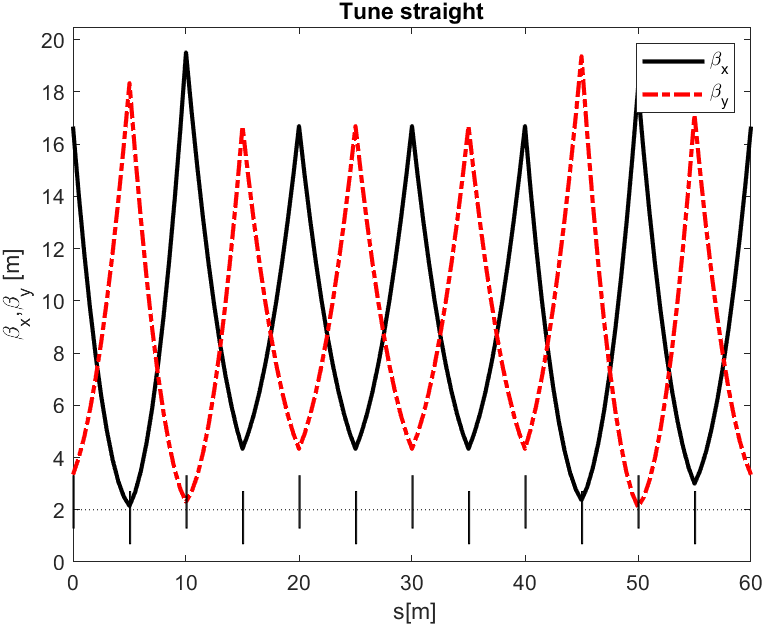}
\end{center}
\caption{\label{fig:tune}Beta functions in the straight section to adjust the tunes.
  The four quadrupoles in the two cells at the ends are used to match the beta functions
  to those of the arcs. The four quadrupoles in the middle are used to adjust the tunes.
}
\end{figure}
%..............................................
The two straight sections consist of six 90-degree cells as shown on the left-hand side in
Figure~\ref{fig:cell}. In one of the straight sections of the racetrack we use two central
cells to adjust the tunes; the two QF and two QD are powered in series. Moreover the two
adjacent cells on either side are used to match the beta functions to those of the arcs.
We adjust the tunes to $26.413$ in the horizontal plane and to $24.528$ in the vertical
plane and show the beta functions in this straight section in Figure~\ref{fig:tune}.
It turned out to be beneficial to have one tune below and the other above the
half-integer, because the skew quadrupoles will push the fractional tunes apart. Conversely,
if both are below the half-integer, one of them will cross the half-integer resonance, which
leads to an unstable lattice.
\par
%..............................................
\begin{figure}[tb]
\begin{center}
\includegraphics[width=0.47\textwidth]{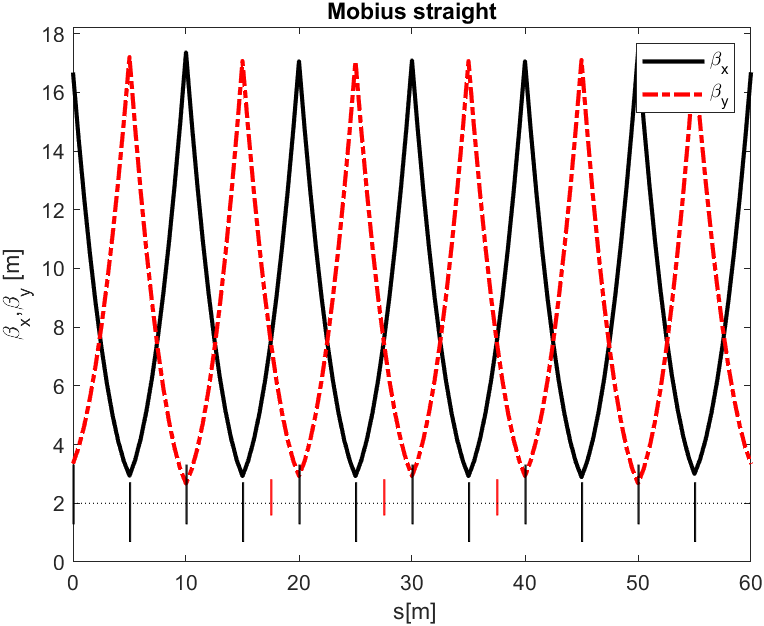}
\includegraphics[width=0.47\textwidth]{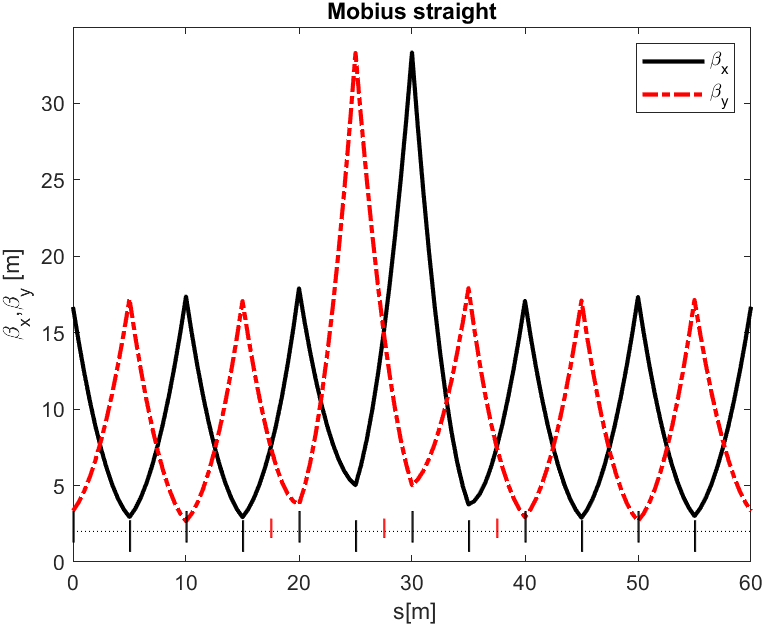}
\end{center}
\caption{\label{fig:mobius}Beta functions in the straight section to adjust the coupling.
  On the left-hand side the skew quadrupoles are turned off and on the right-hand side
  they are turned on. The location of the skew quadrupoles is indicated by the red markers
in the lattice shown near the bottom.}
\end{figure}
% ..............................................
The straight section on the other side of the ring also consists of six 90-degree FODO cells.
With skew quadrupoles turned off, the transfer matrix of the straight section simply matches
one arc to the next. In three of the six cells we place thin-lens skew quadrupoles in the
middle of the drift space between a defocusing and a focusing quadrupole. They are shown by
the red markers in the lattice at the bottom of both plots in Figure~\ref{fig:mobius}. At
this location the value of both beta functions is equal to $\beta_0=3L/2=7.5\,$m and the
phase advance between adjacent skew quadrupoles is $90^o$ in both planes. If the focal length
of the skew quadrupoles $f_s$ is set to $f_s=\beta_0$ the diagonal $2\times2$ blocks of the
transfer matrix for this section are zero, while the off-diagonal blocks are non-zero; it
exchanges the transverse planes. Adjusting the excitations of the three skew quadrupoles by
the same factor therefore allows us to continuously vary the coupling from the uncoupled
configuration shown on the left-hand plot in Figure~\ref{fig:mobius} to the M\"obius
configuration, shown on the right-hand plot.
\section{The ring}
%
%..............................................
\begin{figure}[p]
\begin{center}
  \includegraphics[width=0.98\textwidth]{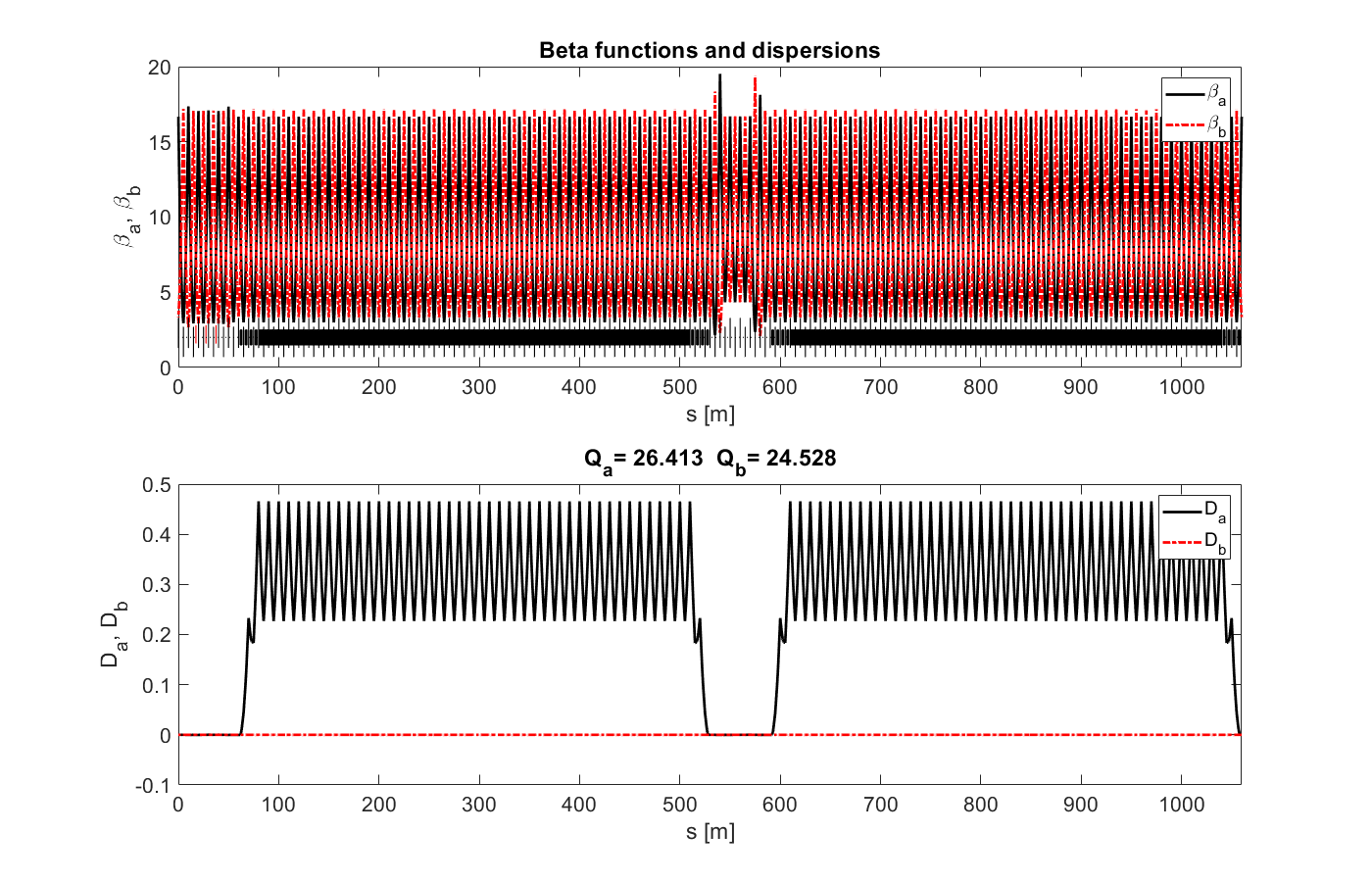} 
  \includegraphics[width=0.81\textwidth]{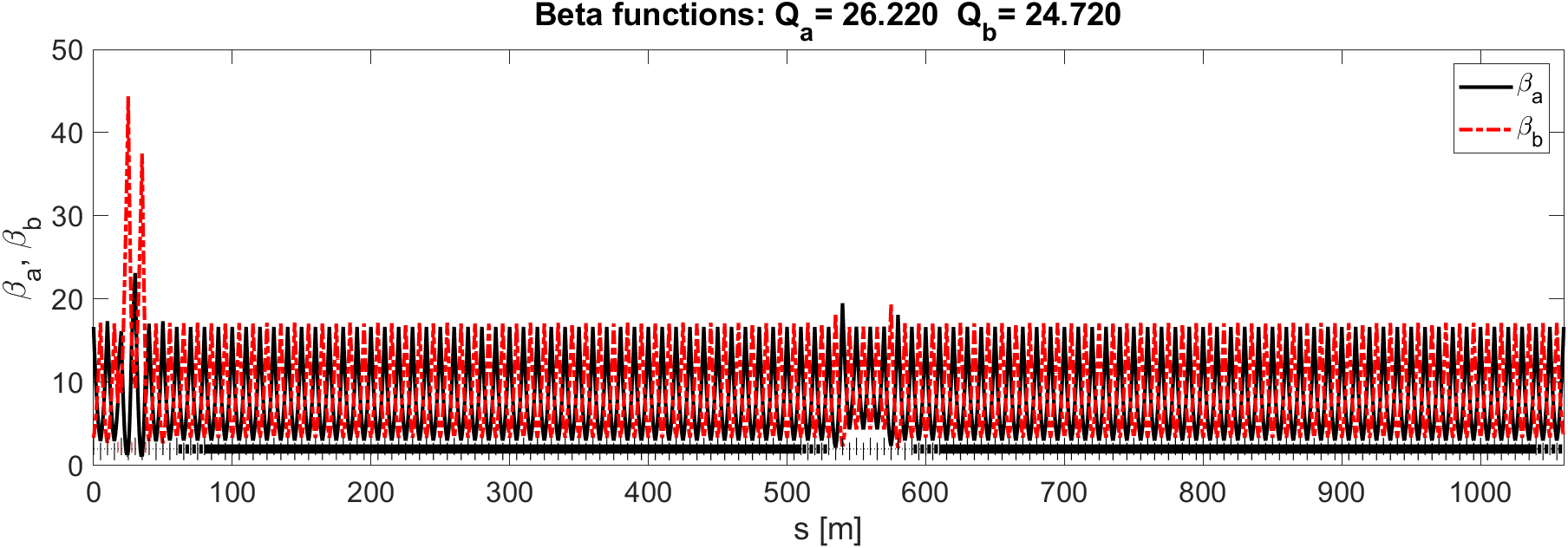}
  \vskip 4mm
  \includegraphics[width=0.81\textwidth]{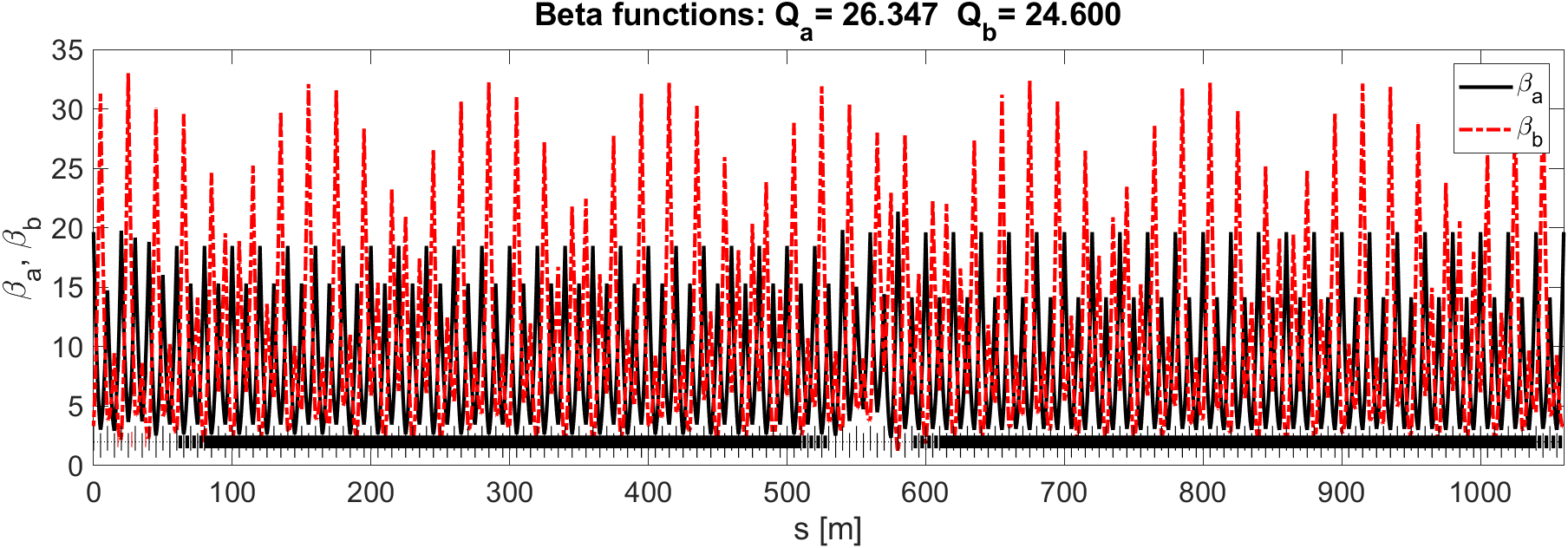}  
\end{center}
\caption{\label{fig:ring}Beta functions and dispersion in the M\"obius ring with
  skew quadrupoles turned off (top). The beta functions with skew quadrupoles turned
  on (middle), and set to a value half way in between (bottom).}
\end{figure}
% ..............................................
The complete ring consists of the two straight sections sandwiched between the arcs
starting with the M\"obius straight. The upper plots in Figure~\ref{fig:ring} show the
beta functions and the dispersions in the uncoupled configuration with skew quadrupoles
turned off. The straight section with skew quadrupoles is located between $s=0$ and
60\,m, where we observe the dispersions to be zero. The straight section to adjust the
tunes is located between $s=500\,$m and 600\,m, where the dispersion also is zero. In
this configuration the tunes are $Q_x=26.413$ and $Q_y=24.528$. 
\par
The middle plot in Figure~\ref{fig:ring} shows the beta functions for the fully-coupled
configuration with focal length of the skew quadrupoles adjusted to $f_s=\beta_0$, which
causes the two $2\times2$ blocks on the diagonal of the transfer matrix for the whole ring
to be zero while the off-diagonal $2\times2$ blocks are non-zero. We calculate the shown
beta functions and tunes in this strongly coupled lattice with the Edwards-Teng
algorithm~\cite{EDTENG}, refined by Sagan and Rubin~\cite{SAGRUB}. In this configuration
the tunes are $Q_a=26.220$ and $Q_b=24.720$. Note that the fractional parts differ by
a half-integer, a feature that we discuss further below. It is noteworthy that,
apart from a moderate mismatch in the straight section with the skew quadrupoles, the
beta functions are equal to those in the uncoupled configuration. A reason is that both
straight sections are matched to the arcs and that the dispersion at the location of the
skew quadrupoles is zero. In this way adjusting the coupling is independent of the
rest of the optics. 
\par
The bottom plot in Figure~\ref{fig:ring} shows the beta functions for a configuration
with focal length of the skew quadrupoles set to $f_s=2\beta_0$, which causes
significant beating of the beta functions in the arcs. We point out that the dispersions
for all three configurations are the same as those shown in the uncoupled configuration.
\par
We now turn to the calculation of the beam parameters as we increase the excitation of the
skew quadrupoles.
\section{Beam parameters}
%
%..............................................
\begin{figure}[tb]
  \begin{center}  
    \includegraphics[width=0.48\textwidth]{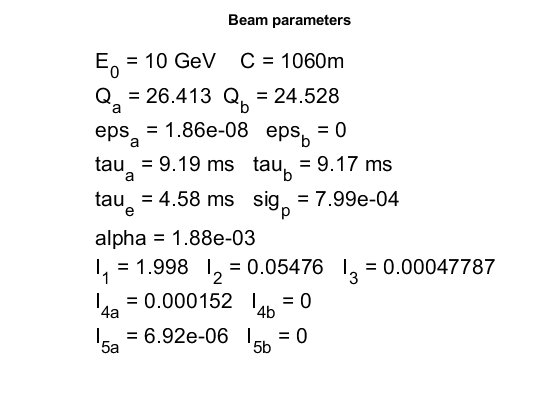}
    \includegraphics[width=0.48\textwidth]{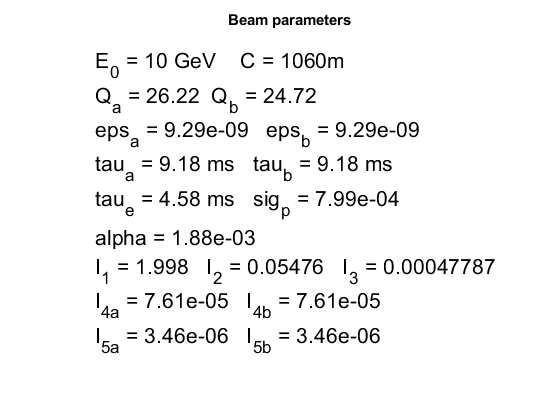}
  \end{center}
  \vskip -10mm
\caption{\label{fig:beampara}Beam parameters for the uncoupled configuration (left) and
  the M\"obius configuration (right).}
\end{figure}
%..............................................
%..............................................
\begin{figure}[p]
  \begin{center}  
    \includegraphics[width=0.8\textwidth]{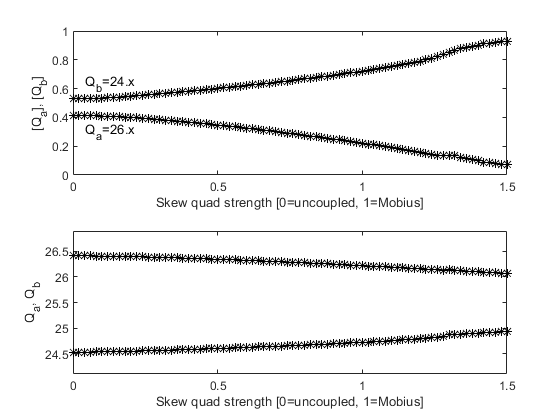}
  \end{center}
  \caption{\label{fig:tunes}The fractional tunes (top) and the full tunes (bottom)
    as a function of the excitation of the skew quadrupoles.}
  \begin{center}  
    \includegraphics[width=0.8\textwidth]{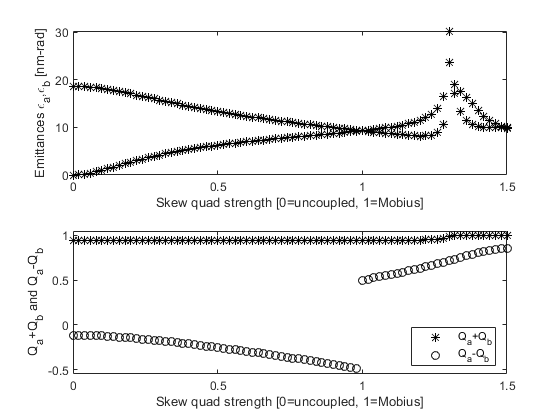}
  \end{center}
  \caption{\label{fig:emit}The emittances (top) and the difference and sum of the
    fractional tunes (bottom) as a function of the excitation of the skew quadrupoles.}
\end{figure}
%..............................................
We now calculate the beam parameters using the algorithm described in~\cite{SZ}. Those
corresponding to the configurations for the uncoupled and the M\"obius
ring that gave rise to the upper and middle plots in Figure~\ref{fig:ring} are shown in
Figure~\ref{fig:beampara}. We observe that the emittances for the uncoupled lattice
are 18.6\,nm-rad and zero, whereas for the M\"obius configuration both emittances are
9.3\,nm-rad. We also find that the small differences in the transverse damping times
of the uncoupled configuration vanishes in the M\"obius configuration. The longitudinal
parameters for both configurations are equal. In the rows below we
show the radiation integrals. Only $I_{4a}$ and $I_{4b}$ as well as
$I_{5a}$ and $I_{5b}$ differ between the configurations, where the corresponding values
in the M\"obius configuration are equal to half the horizontal values $I_{4a}$ and
$I_{5a}$ of the uncoupled configuration. In the M\"obius configuration, the effect of
damping and excitation is shared equally between the two planes. 
\par
Since only the tunes and emittances show a significant difference between the two
configurations, we explore their variation further as we increase the strength of
the skew quadrupoles. On the upper panel in Figure~\ref{fig:tunes} we show the change
of the fractional tunes. We observe that the increasing skew quadrupoles ``push the
fractional tunes apart.'' Note, however, that the upper branch corresponds to the
``vertical'' tune $Q_b$ and the lower branch to the ``horizontal'' tune $Q_a.$
The lower panel also shows both tunes but including their integral part.
\par
The upper panel in Figure~\ref{fig:emit} shows the emittances as a function of the
excitation of the skew quadrupoles. We see that initially $\eps_a$ is non-zero
and $\eps_b$ is zero, but as the coupling increases their values get closer and
become equal in the M\"obius configuration. This is accompanied by the difference
of the fractional tunes approaching a half-integer. This is apparent from the
circles on the lower panel, which show $Q_a-Q_b$. If we increase the excitation
beyond the M\"obius configuration both emittances increase dramatically. The plot
on the lower panel provides an explanation; the sum of the tunes $Q_a+Q_b$ becomes
an integer and the system crosses a sum resonance shown by the asterisks in the
lower panel from Figure~\ref{fig:emit}. And on a sum resonance, the emittances can
become arbitrarily large, because only their difference is bounded~\cite{SUMRESO,SUMHB}.
\section{Conclusions}
In order to explore the ability of the new method to calculate synchrotron
radiation integrals~\cite{SZ} to handle highly-coupled systems we designed a
ring, based on 90-degree FODO cells, which can be continuously tuned from an
uncoupled to a M\"obius configuration. The method works well and allow us
to calculate all beam parameters that depend on the synchrotron radiation
integrals. In particular, the damping times only show a small dependence
on the coupling whereas the emittances, which differ greatly in the uncoupled
configuration, become equal in the M\"obius configuration. If we increase the
coupling beyond the M\"obius configuration, the system encounters a sum resonance,
where both emittances grow significantly, consistent with~\cite{SUMRESO,SUMHB}.
\par
Having the emittances and momentum spread available, we can calculate the
self-consistent equilibrium beam sizes in all three dimensions, even for
highly-coupled lattices.
\par
Apart from serving as a test ground for the new method to calculate synchrotron
radiation integrals, the M\"obius ring might become useful to explore the
effect of coupling on orbit correction, non-linear effects, and the beam-beam
interaction.
%
%%%%%%%%%%%%%%%%%%%%%%%%%%%%%%%%%%%%%%%%%%%%%%%%%%%%%%%%%%%%%%%%%%%%%%%
%
\bibliographystyle{plain}

\begin{thebibliography}{M}
  %
\bibitem{SZ}
  V. Ziemann, A. Streun, {\em Equilibrium parameters in coupled storage ring lattices and
    practical applications}, in preparation.
\bibitem{HELM}
  Richard H. Helm, Martin J. Lee, P. L. Morton, and M. Sands, {\em Evaluation of synchrotron
    radiation integrals}, IEEE Trans. Nucl. Sci. 20 (1973) 900.
\bibitem{TALMAN}
  R. Talman, {\em A proposed M\"obius accelerator,} Physical Review Letters 74 (1995) 1590.
\bibitem{SUPMAT}
  Software used to design the M\"obius ring is available from
  \url{https://github.com/volkziem/MobiusRing}.
\bibitem{VZAP}
  V. Ziemann, {\em Hands-On Accelerator Physics Using MATLAB,} CRC Press, Boca Raton,
  2019.
\bibitem{AIBA}
  M. Aiba, M. Ehrlichman, A. Streun, {\em Round beam operation in electron storage rings
    and generalization of M\"obius accelerator,} Proceedings of IPAC 2015 in Richmond VA,
  USA, (2015) 1716.
\bibitem{EDTENG}
  D. Edwards, L. Teng, {\em Parameterization of linear coupled motion in periodic systems,}
  IEEE Trans. Nucl. Sci. 20 (1973) 885.
\bibitem{SAGRUB}
 D. Sagan and D. Rubin, {\em Linear analysis of coupled lattices,} Phys. Rev. ST Accel. Beams,
 2:074001, 1999. [Phys. Rev. ST Accel. Beams3,059901(2000)].
\bibitem{SUMRESO}
  Section 1.3.1 in H. Wiedemann, {\em Particle Accelerator Physics II, 2nd ed.,} Springer Verlag,
  Heidelberg, 2003.
\bibitem{SUMHB}
  Section 2.1.3 in A. Chao, M. Tigner, {\em Handbook of Accelerator Physics and Engineering,} World Scientific,
  Singapore, 1999.
% %
\end{thebibliography}

\end{document}